# BN/Graphene/BN Transistors for RF Applications

Han Wang[1], Thiti Taychatanapat[1], Allen Hsu[1], Kenji Watanabe[2], Takashi Taniguchi[2], Pablo Jarillo-Herrero[1], and Tomas Palacios[1]

*Abstract*— In this letter, we demonstrate the first BN/Graphene/BN field effect transistor for RF applications. The BN/Graphene/BN structure can preserve the high mobility of graphene, even when it is sandwiched between a substrate and a gate dielectric. Field effect transistors (FETs) using a bilayer graphene channel have been fabricated with a gate length $L_G$=450 nm. A current density in excess of 1 A/mm and DC transconductance close to 250 mS/mm are achieved for both electron and hole conductions. RF characterization is performed for the first time on this device structure, giving a current-gain cut-off frequency $f_T$=33 GHz and an $f_T.L_G$ product of 15 GHz.μm. The improved performance obtained by the BN/Graphene/BN structure is very promising to enable the next generation of high frequency graphene RF electronics.

## I. INTRODUCTION

Graphene is a one-atom-thick layer of carbon atoms arranged in a honeycomb lattice through $sp^2$ bonding [1]. Considered for many years an impossible goal, the isolation of graphene triggered a revolution not only among condensed-matter physicists but also among chemists and engineers, eager to take advantage of its unique properties [2]. The advantages of graphene for radio-frequency (RF) applications derive in part from its high electron and hole mobility, which can exceed 100,000 $cm^2$/V.s at T=240 K [3]. In addition, the combination of the unique properties of this material, with new device concepts and nanotechnology may overcome some of the main limitations of traditional RF electronics in terms of maximum frequency, linearity and power dissipation [4, 5]. Recently, high frequency graphene field effect transistors (GFET) have been demonstrated by several groups [6, 7]. However, despite the excellent RF performance achieved, these devices have carrier mobilities below 2,000 $cm^2$/V.s, which are mainly limited by the interaction of graphene with the substrate and gate dielectric.

Most of today's RF GFETs are fabricated on either $SiO_2$ [7] or SiC [6] substrate. Graphene was first isolated on $SiO_2$ due to the ability to identify single layer graphene using optical microscopes while the growth of graphene on SiC provides a natural substrate for these devices. However, neither $SiO_2$ nor SiC are ideal substrates for graphene. One problem with thermally grown $SiO_2$ (a few hundred nm thick) is that it often leads to a high surface roughness, as shown in Figure 1(a). In addition, the oxide typically has a large density of charge traps and defects. Graphene on SiC, on the other hand, suffers from a terraced rough substrate surface that can limit device performance by scattering charge carriers flowing in the active graphene layer [8]. Hence, in order to take full advantage of the ultra high mobility promised by graphene, we need to either remove the substrate [3] or use a better one. Although suspended graphene sheets have shown the highest mobility ever measured at room temperature in any semiconductor, the fragile suspended graphene membrane leads to many fabrication challenges and reliability issues. An alternative approach is to use a better substrate, such as hexagonal boron nitride (hBN) [9-11], which has the same atomic structure as graphene and shares many of its properties. The 2D planar structure of hBN also gives this material an ultra flat surface (Figure 1(a)) that is also free of dangling bonds and charge traps. Hence, it provides an ideal environment for graphene to sit on. Recent work has shown carrier mobility as high as 40,000 $cm^2$/V.s in bilayer graphene (BLG) on hBN at room temperature [9].

In this work, we demonstrate the first BN/Graphene/BN RF field effect transistor, which has hBN as both the substrate and the gate dielectric with bilayer graphene as the channel material. This novel structure can preserve the high carrier mobility in the bilayer graphene channel and hence has a great potential for high frequency transistor applications.

## II. BN/GRAPHENE/BN FETS

The fabrication process of the BN/Graphene/BN devices studied in this work is summarized in Figure 2(a-d). A hexagonal boron nitride flake is first exfoliated on a $SiO_2$/Si substrate. A separate $SiO_2$/Si sample is then coated with polyvinyl acetates (PVA) and polymethyl methacrylate (PMMA); and bilayer graphene flakes from natural graphite are exfoliated on top of the PMMA and transferred using the technique described in Ref. [10]. A flip chip bonder is used in the transfer process to allow for an accurate alignment between the hBN flake and the bilayer graphene flake. The alignment accuracy is within 1-2 μm. Figure 2(e) shows an optical micrograph of a bilayer graphene flake transferred on top of an

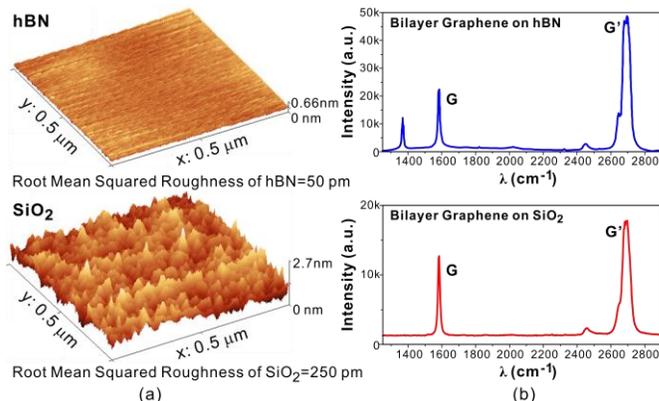

Fig. 1 (a) AFM images showing the surface roughness of hBN and 285 nm thermally grown $SiO_2$. (b) Raman spectra of bilayer graphene on hBN (the peak at λ=1375 $cm^{-1}$ is due to hBN substrate); and bilayer graphene on $SiO_2$.

This work was partially supported by the Army Research Laboratory and the ONR GATE MURI program. P. J-H. also acknowledges support by NSF Career Award (no. 0845287).

[1]Massachusetts Institute of Technology, Cambridge, MA 02139 USA (e-mail: hanw@mtl.mit.edu).

[2]National Institute for Materials Science, Namiki 1-1, Tsukuba, Ibaraki 305-0044, Japan

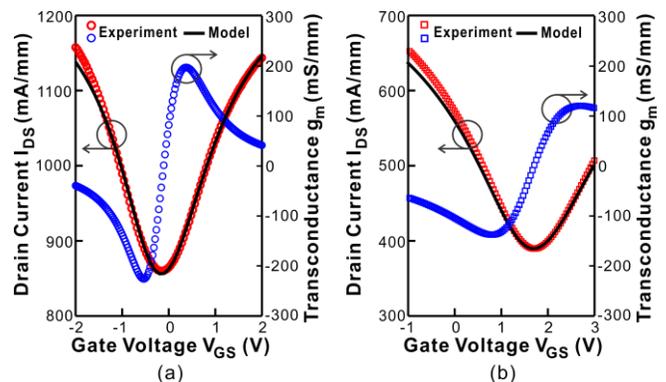

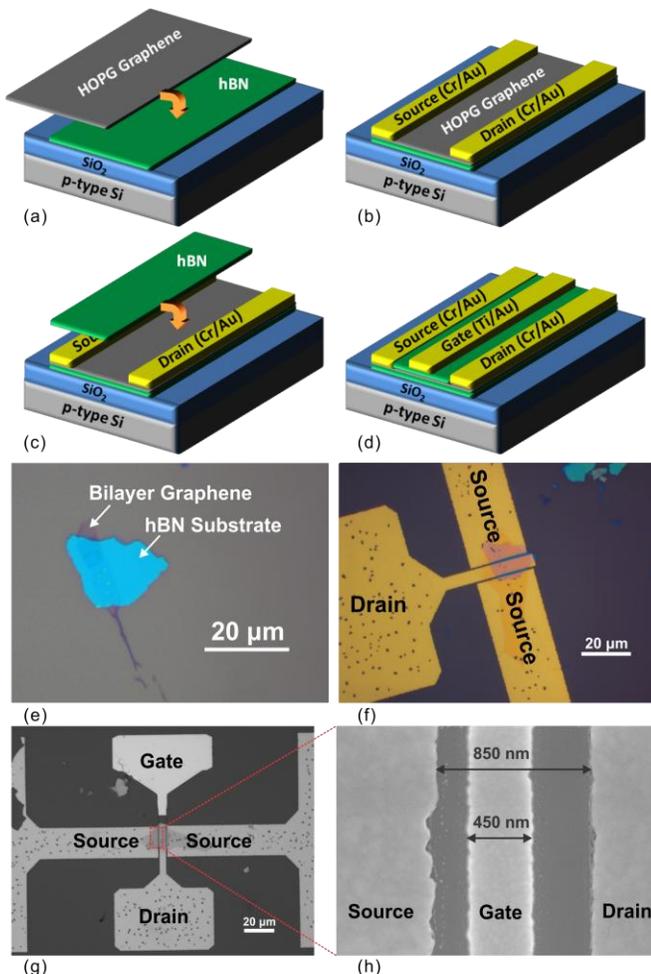

Fig. 2 (a)-(d) Fabrication process for making BN/Graphene/BN FETs. (e) optical micrograph showing a bilayer graphene flake exfoliated and transferred on top of hBN substrate. (f) optical micrograph of the device after source and drain are defined by e-beam lithography and the second (top) layer of hBN has been transferred on to graphene. (g) optical micrograph and (h) SEM image showing the final device. $L_G$=450 nm. $L_{DS}$=850 nm.

hBN substrate. Then, the source and drain contacts are defined by e-beam lithography and formed by depositing a 0.7 nm Cr /50 nm Au metal stack using thermal evaporation. The contact resistance is typically 200 Ω.μm. After that, a second layer of hBN is exfoliated, aligned and transferred on top of the bilayer graphene channel. This second hBN flake becomes the top gate dielectric of the GFET. Figure 2(f) shows the device after the source and drain contacts are made and the top hBN layer has been transferred. Finally, the top gate of the device is defined by e-beam lithography and formed by a 10 nm Ti/40 nm Au multilayer using e-beam evaporation. Figures 2(g) and 2(f) show the finished device, which has a gate length $L_G$=450 nm and source-drain distance $L_{DS}$=850 nm. The width of the device active region is 3 μm. The top gate dielectric has a thickness of 8.6 nm as measured by atomic force microscopy (AFM), which gives a top-gate capacitance of 0.39 μF/cm$^2$. The conductive Si substrate forms the bottom gate of the device with a dielectric consisting of 285 nm SiO$_2$ plus the hBN substrate layer. The top and bottom gate capacitances ratio is $C_{TG}/C_{BG} \approx 30$.

## III. RESULTS AND DISCUSSIONS

Figure 3(a) shows the transfer characteristics ($I_{DS}$ vs. $V_{GS}$) of the fabricated transistor. At $V_{DS}$=1 V, the device achieves a high current density close to 1.2 A/mm. The minimum conduction point of the transistor is very close to 0 V. This indicates negligible doping effects from both the hBN substrate and the hBN top gate dielectric. Typically, bilayer graphene on hBN substrate produced in our laboratory show Hall mobilities in excess of 15,000 cm$^2$/V.s. In addition, there is very little mobility degradation after the second hBN layer (i.e. the top gate dielectric) is transferred on top. The maximum transconductance $g_m$ is close to 250 mS/mm. The extrinsic DC transconductance is mainly limited by the access resistances and gate capacitance, not the mobility.

Fig. 3 DC characteristics with the substrate grounded of (a) the BN/Graphene/BN device, and (b) the control device. $V_{DS}$=1 V in both devices. In both (a) and (b), the black solid lines show the fit by the model proposed in [12] to the device transfer characteristics.

Figure 4(a) shows the RF performance of the device. With $L_G$=450 nm and at $V_{DS}$=1 V, the device has a current-gain cut-off frequency $f_T$=5 GHz and $f_T$=22 GHz before and after de-embedding the coplanar-waveguide (CPW) pad capacitances, respectively. The de-embedding procedure follows the well-established standard open-short method [13, 6, 7]. In these measurements, the back-gate, i.e. substrate, was grounded.

The substrate bias has a significant effect on the RF performance of the device. With the substrate grounded, the un-gated access regions on both the source and the drain side of the device have their Fermi energy levels located near the Dirac point (the minimum conduction point). This leads to relatively large source and drain access resistances ($R_S$ and $R_D$). The resistances of these un-gated regions can be reduced through electrostatically doping them by biasing the substrate. This reduction in the resistances significantly increases the frequency performance of the device [14]. For example, Figure 4(b) shows that when the substrate is biased at -30 V, the current gain cut-off frequency increases to 6 GHz and 33 GHz before and after de-embedding.

The DC and RF performance of this device was also compared to a control device fabricated on a SiO$_2$ substrate and with a 16 nm Al$_2$O$_3$ gate dielectric. This control device also uses a bilayer graphene flake exfoliated from natural graphite as the channel material. The Al$_2$O$_3$ gate dielectric is formed by naturally oxidizing 3 nm of e-beam evaporated Al followed by atomic layer deposition of 13 nm Al$_2$O$_3$. The thickness of the Al$_2$O$_3$ gate dielectric was chosen to render the same top gate capacitance as in the BN/Graphene/BN FET. The Hall mobility of our bilayer graphene on SiO$_2$ is typically between 1,500 and 2,000 cm$^2$/V.s [15]. This low mobility is mainly due to the scattering introduced by the SiO$_2$ substrate. The mobility

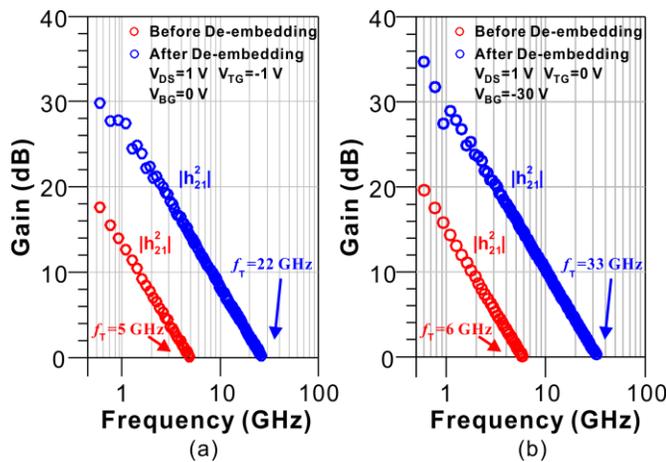

(a)    (b)

Fig. 4 RF performance of the fabricated BN/Graphene/BN FET with $L_G$=450 nm when (a) the substrate is biased at $V_{BG}$=0 V, and (b) the substrate is biased at $V_{BG}$=-30 V. In both (a) and (b), the $|h_{21}^2|$ data and $f_T$ before de-embedding are shown in red; and that after de-embedding are shown in blue. $V_{DS}$=1 V. The top gate bias is $V_{TG}$=-1 V for (a) and $V_{TG}$=0 V for (b), such that the device is very close to the maximum $g_m$ in each case.

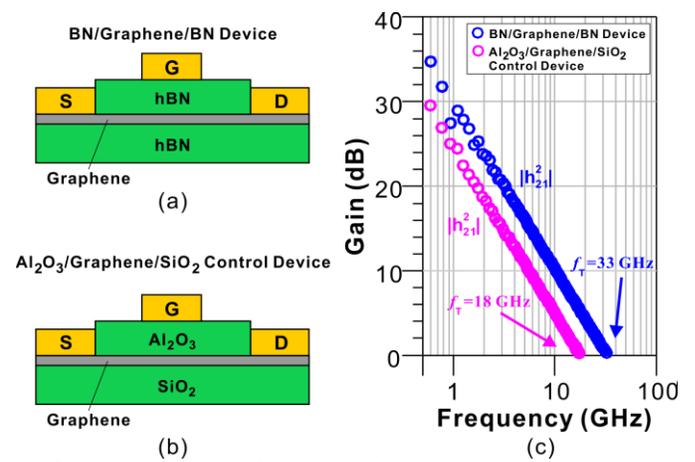

(b)    (c)

Fig. 5 (a) Schematic of the BN/Graphene/BN device structure. (b) Schematic of the $Al_2O_3$/Graphene/$SiO_2$ control device structure. (c) Comparison of BN/Graphene/BN device with the control device in terms maximum $f_T$.

degrades by ~30% after oxide deposition due to the additional impurities introduce to the system.

For similar device dimension and gate capacitances, the BN/Graphene/BN FET shows a peak transconductance ($g_m$) of 250 mS/mm, which is about 70% higher than that in the control sample that has a peak $g_m$ of only 140 mS/mm (Figure 3). To further analyze and compare the two devices, the virtual source model proposed in Ref. [12] is used to fit their DC characteristics (Figure 3). The model extracts a field effect mobility of 6,500 $cm^2$/V.s in the BN/Graphene/BN FET and 1,200 $cm^2$/V.s in the control device, respectively. This carrier mobility, while much higher than the control device but relatively low compared to other reported mobility values of bilayer graphene on hBN, is possibly due to bubbles and ripples created during the transfer process; and the measurements being taken at high drain biases ($V_{DS}$=1 V) and high current density. The carrier injection velocities are estimated to be about $3.5\times10^7$ cm/s in the BN/Graphene/BN FET and $2.5\times10^7$ cm/s in the control device. This gives an indication of the significant advantage that an hBN substrate and dielectric can have over $SiO_2$ and $Al_2O_3$ in terms of preserving the high carrier mobility and carrier velocity in graphene. Figure 5 compares the peak $f_T$ of these two devices of equal gate length and gate capacitance. For $V_{DS}$=1 V, the BN/Graphene/BN FET has its highest $f_T$=33 GHz at $V_{BG}$=-30 V, and $V_{TG}$=0 V while the control sample has its highest $f_T$=18 GHz at $V_{BG}$=-10 V, and $V_{TG}$=1 V, demonstrating a significant improvement in peak $f_T$ due to the change of substrate and the gate dielectric material.

## IV. Conclusion

In this letter, we fabricated the first BN/Graphene/BN RF FET and characterized its DC and RF performances. This new device is also compared to a control GFET with a $SiO_2$ substrate and an $Al_2O_3$ gate dielectric. The BN/Graphene/BN structure allows a much higher mobility and carrier velocity than in the case of $SiO_2$ substrates and $Al_2O_3$ gate dielectrics. With the same device dimensions, the BN/Graphene/BN device shows a significant improvement in $f_T$ compared to the control device, demonstrating its great potential for applications in high frequency electronic circuits. In addition, recent developments in the synthesis of both graphene [16] and hBN [17] by chemical vapor deposition (CVD) method may allow the technology proposed in this paper to be implemented at wafer-scale in the near future.

### Acknowledgements
The authors would like to thank Javier Sanchez-Yamagishi for his help with the experiments.